\documentclass[twocolumn]{aastex62}
\usepackage{natbib}
\usepackage{amsmath}
\usepackage{bm}
\usepackage{graphicx}
\usepackage{subfigure}
\usepackage{float}
\usepackage{epsfig}
\usepackage{booktabs}
\usepackage[utf8x]{inputenc}
\usepackage[encapsulated]{CJK}
\usepackage{physics}
\usepackage{soul}

\newcommand{\cntextsc}[1]{\begin{CJK*}{UTF8}{gbsn}#1\end{CJK*}}
\newcommand{\jptext}[1]{\begin{CJK}{UTF8}{min}#1\end{CJK}}

\graphicspath{{./}{figures/}}

\shorttitle{Non-ideal MHD simulation of HL Tau disk}
\shortauthors{Hu et al.}

\begin{document}
\title{Non-ideal MHD simulation of HL Tau disk: formation of rings}

\author{Xiao Hu (\cntextsc{胡晓})}
\affil{Department of Physics and Astronomy, University of
Nevada, Las Vegas, 4505 South Maryland Parkway, Las Vegas,
NV 89154, USA}
\author{Zhaohuan Zhu (\cntextsc{朱照寰})}
\affil{Department of Physics and Astronomy, University of
Nevada, Las Vegas, 4505 South Maryland Parkway, Las Vegas,
NV 89154, USA}
\author{Satoshi Okuzumi (\jptext{奥住聡})}
\affil{Department of Earth and Planetary Sciences, Tokyo Institute of Technology, Meguro, Tokyo, 152-8551, Japan}
\author{Xue-Ning Bai (\cntextsc{白雪宁})}
\affil{Institute for Advanced Study, Tsinghua University, Beijing 100084, China}
\affil{Department of Astronomy, Tsinghua University, Beijing 100084, China}
\author{Lile Wang (\cntextsc{王力乐})}
\affil{Center for Computational Astrophysics, Flatiron Institute, New York, NY 10010, USA}
\author{Kengo Tomida (\jptext{富田 賢吾})}
\affil{Department of Earth and Space Science, Osaka University, Toyonaka, Osaka, 560-0043, Japan}
\author{James M. Stone}
\affil{Department of Astrophysical Sciences, 4 Ivy Lane, Peyton
Hall, Princeton University, Princeton, NJ 08544, USA}

\begin{abstract}
Recent high resolution observations unveil ring structures in circumstellar disks. The origin of these rings has been widely investigated under various theoretical scenarios. In this work we perform global 3D non-ideal MHD simulations including effects from both Ohmic resistivity and ambipolar diffusion (AD) to model the HL Tau disk. The non-ideal MHD diffusion profiles are calculated based on the global dust evolution calculation including sintering effects.  Disk ionization structure changes dramatically across the snow line due to the change of dust size distribution close to snow line of major volatiles. We find that accretion is mainly driven by disk wind. Gaps and rings can be quickly produced from different accretion rates across snow line. Furthermore, ambipolar diffusion (AD) leads to highly preferential accretion at midplane, followed by magnetic reconnection. This results a local zone of decretion that drains of mass in the field reconnection area, which leaves a gap and an adjacent ring just outside it. Overall, under the favorable condition, both snow lines and non-ideal MHD effects can lead to gaseous gaps and rings in protoplanetary disks.
\end{abstract}


\section{Introduction} \label{sec:intro}

Accretion disks play an important role in forming many astrophysical objects. Protoplanetary disks, while supplying mass to central stars, are also birth places of planets. HL Tau is a Class I-II protostellar source surrounded by a circumstellar disk with a size of $\sim$100 au. The 2014 Long Baseline Campaign of the Atacama Large Millimeter/submillimeter Array (ALMA) provided unprecedented details of the HL Tau disk\citep{2015ApJ...808L...3A}.
Patterns of multiple bright and dark rings that are symmetric to the central
star are clearly seen. Subsequent ALMA observations find rings in dust continuum emission are very common in protoplanetary disks \citep[e.g.,][]{2018ApJ...869L..42H,2018ApJ...869...17L}.

Axisymmetric dust rings in a protoplanetary disk can be produced in several ways. With dust radial drift under the action of gas drag, a major mechanism of creating dust radial concentration is dust trapping at local gas pressure maxima \citep{1972fpp..conf..211W}. Planet-disk interaction has been the most popular way to carve gaps in protoplanetary disks. The filtration effect on larger dust particles can make dust gaps more prominent than gas structure \citep[e.g.,][]{2012ApJ...755....6Z,2018ApJ...869L..47Z}. The gap feature in HL Tau can be either explained by multi planets with one of them in each gap\citep[e.g.,][]{2015ApJ...809...93D,2015MNRAS.453L..73D}, or a single planet opening multiple gaps\citep{2018ApJ...864L..26B}.

Another possibility is the condensation fronts of varies solids altering dust growth. The large dispersion of location and size of rings in from recent observations show low correlation to snow lines or resonance orbits of planets \citep{2018ApJ...869L..42H,2018ApJ...869...17L,2019ApJ...872..112V}. Though snow lines are not likely to be the common origin of rings and gaps, it still does a good job in the case of HL Tau. It was first proposed by \citet{2015ApJ...806L...7Z} that smaller dust can be depleted by rapid pebble growth at snow lines. The location of dark rings in the ALMA image are close to snow lines of some major volatiles, e.g.,$\mathrm{H_2O}$ and $\mathrm{NH_3}$. A more detailed study on dust coagulation by \citet{2016ApJ...821...82O} shows sintering effect between dust aggregates can change dust sizes near snow lines even with low volatile abundance. 

Other theoretical models include secular gravitation instability triggered from low turbulence inner disk making ring-like structure \citep{2016AJ....152..184T}. Dust to gas back reaction and gas compression can cause large gradient of dust radial drift velocity, this self induced dust pile-ups is discussed in \citet{2017MNRAS.467.1984G}. If planets are formed early, dust rings can be the collision outcome of planetesimals perturbed by planets and orbital resonances \citep{2017ApJ...850..103B}. Magnetic self-organization is seen in form of `zonal flows', resulting radial variations in density and magnetic flux. These banded rings can sustain as the pressure gradient is balanced by the Coriolis force \citep{2009ApJ...697.1269J,2014ApJ...796...31B}. Recent MHD simulations suggest formation of rings and gaps can be achieved through redistribution of magnetic flux, direct feeding of avalanche accretion and midplane magnetic reconnection\citep{2017MNRAS.468.3850S,2018MNRAS.477.1239S, 2019MNRAS.484..107S}, when non-ideal MHD diffusivities like Ohmic resistivity and ambipolar diffusion are considered.

The abundance of dust is not just a disk profile tracer in radio continuum observation, but can also play a crucial role constraining disk accretion. In the traditional scenario of viscous driven accretion, gas surface density varies at radial viscosity jumps such as transition from magnetorotational instability (MRI) active zone to dead zone \citep[e.g.,][]{1996ApJ...457..355G,2015A&A...574A..68F,2016ApJ...816...19H}.  Disk ionization degree strongly depends on the abundance of small (sub-micron) grains, thus affecting the coupling between gas and magnetic fields \mbox{\citep[e.g.,][]{2007Ap&SS.311...35W,2011ApJ...739...51B}}. This inspired the idea, that will be presented in this paper, to incorporate snow line induced dust distribution into MHD disk simulations. In the sintering operating regime, easier fragmentation produces more smaller grains \citep{2016ApJ...821...82O}, reduces ionization rate and increases magnetic diffusion. This could generate radial variantion of acccretion rate, leads to axisymmetric structure in gas surface density.

In this study we first generated non-ideal MHD diffusivity profile from dust coagulation model including sintering effect. Then we implement the diffusion profile in global 3D non ideal MHD simulation to analyze disk accretion and density structure. In Section \ref{sec:numerical} we briefly introduce basic physics and numerical setups. In Section \ref{sec:results} we present diagnostics of disk radial and vertical structure. In Section \ref{sec:parameter} we explore the impact of variation field strength, shape and magnetic diffusion. We will summarize the paper in section \ref{sec:sum}.

\section{Numerical setup}\label{sec:numerical}
\subsection{Non ideal MHD diffusion}
We solve the magnetohydrodynamic (MHD) equations using Athena++ (Stone et al.
2019, in preparation) with non-ideal magnetic diffusion terms \citep{2017ApJ...836...46B,2017ApJ...845...75B}. The basic non-ideal induction equations are:
\begin{eqnarray}
\frac{\partial {\bm B}}{\partial t}=\nabla \times \left({\bm v}\times {\bm B}\right)
-\frac{4\pi}{c}\nabla \times \left( \eta_\mathrm{O} {\bm J} + \eta_\mathrm{H} {\bm J}\times {\bm b} + 
\eta_\mathrm{A} {\bm J}_{\bot}\right),\nonumber \\ 
\label{eq:induction}
\end{eqnarray}
where $\bm v$ is gas velocity, $\bm B$ is magnetic field, ${\bm b} =
{\bm B}/|B|$ is the unit vector representing field line direction.  
$\bm J$ is current density vector, with $\bm J_{\bot}$ as the 
current component perpendicular to the magnetic field. $\eta_\mathrm{O}$,
$\eta_\mathrm{H}$ and $\eta_\mathrm{A}$ are Ohmic, Hall and ambipolar diffusivities, and
here for simplicity, we only consider Ohmic resistivity and ambibolar diffusion (AD). The Hall diffusivity significantly complicates the picture, but they are more applicable to the inner disk
\mbox{\citep{2016A&A...589A..87B,2017ApJ...845...75B}}. Modeling HL Tau does not to resolve the inner disk,so our simulation mainly focused on outer part disk where most snow lines are. 

\begin{figure}
\centering
\includegraphics[width=0.45\textwidth]{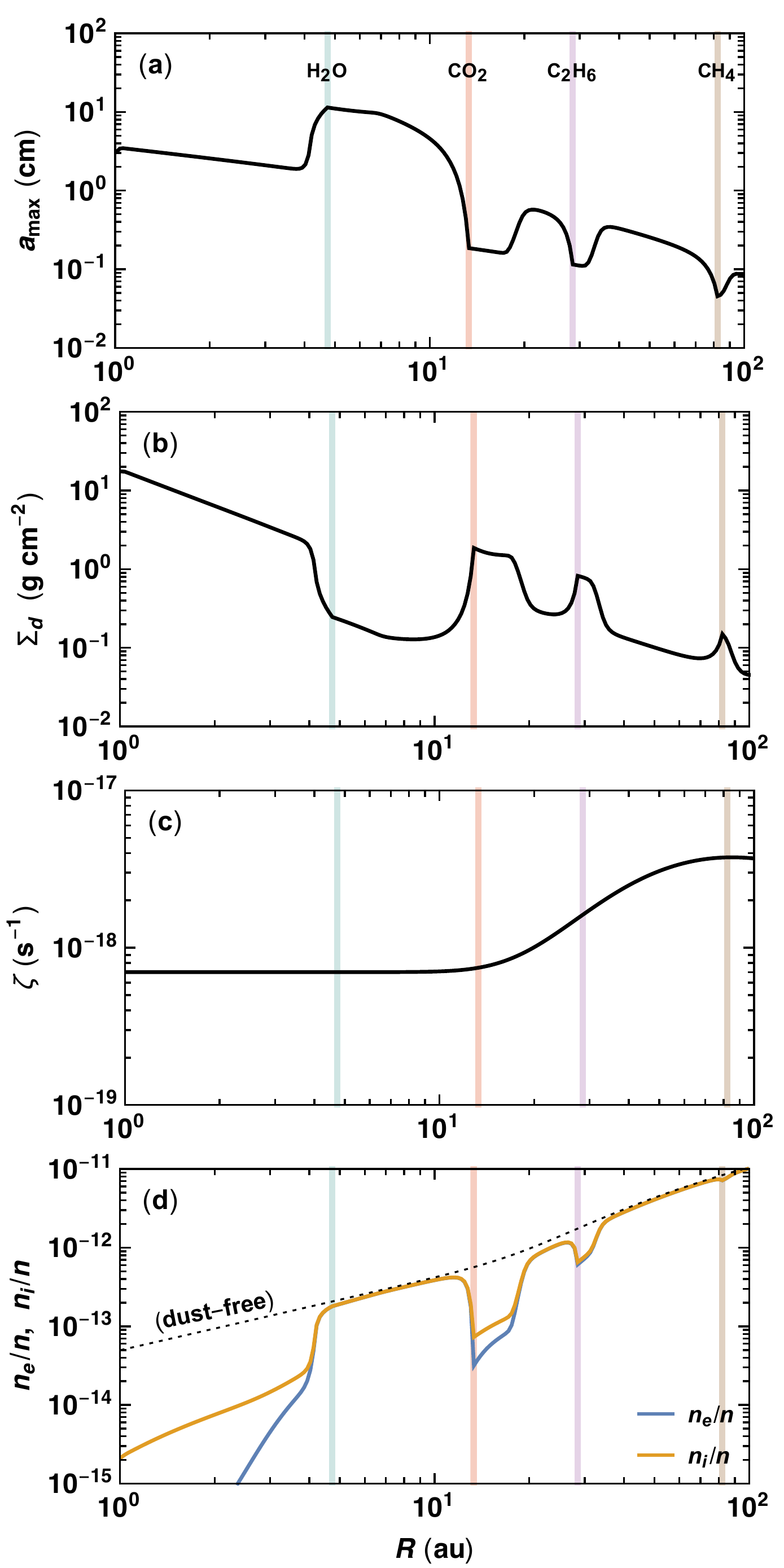}
\caption{
Radial profiles of the maximum grain size $a_{\rm min}$ (panel (a)), total dust surface density $\Sigma_d$ (panel (b)), and midplane ionization rate $\zeta$ (panel (c)) used in the ionization calculation,
as well as the profiles 
of the electron and ion abundances, $n_e/n$ and $n_i/n$, obtained from the ionization calculation (panel (d)).
The vertical lines mark the snow lines of, from left to right,  H$_2$O, CO$_2$, C$_2$H$_6$, and CH$_4$. For comparison, the ion and electron abundances in the dust-free case are also shown by the dotted curve in panel (d).}
\label{fig:dust}
\centering
\end{figure}
We compute the radial and vertical profiles of the magnetic diffusivities in three steps. The first step is to calculate the radial ionization structure at the midplane by considering the balance among ionization, recombination in the gas, and grain charging.
The grain size distribution is given by a power law with minimum and maximum grain sizes $a_{\rm min}$ and $a_{\rm max}$ (Equation~20 of \citealt{2016ApJ...821...82O}). The minimum grain size $a_{\rm min}$ is fixed to $0.1~\micron$, whereas the maximum size $a_{\rm max}$ and  total dust surface density $\Sigma_d$ are taken from a snapshot of a dust evolution calculation that takes into account the low stickiness of CO$_2$ ice \citep{2019ApJ...878..132O} and aggregate sintering \citep{2016ApJ...821...82O}.
 The fragmentation threshold velocity of nonsintered aggregates is assumed to be $0.73~\rm m~s^{-1}$ inside the H$_2$O snow line, $7.3~\rm m~s^{-1}$ between the H$_2$O and CO$_2$ snow lines, and $0.73~\rm m~s^{-1}$ outside the CO$_2$ snow line. In regions where aggregate sintering takes place, the fragmentation threshold is taken to be $40\%$ of the threshold for nonsintered aggregates. The dust evolution calculation assumes weak gas turbulence with a velocity dispersion of 1\% of the sound speed.
The snapshot is taken at the time at which the computed total millimeter fluxes from the dust disk match the ALMA observations \citep{2015ApJ...808L...3A}.
The adopted radial profiles of $a_{\rm max}$ and $\Sigma_d$ are plotted in Figures~\ref{fig:dust}(a) and (b). Because of sintering, dust particles behind the snow lines CO$_2$, C$_2$H$_6$, and CH$_4$ experience enhanced collisional fragmentation and pile up there. The loss of H$_2$O ice interior to the H$_2$O snow line also enhances collisional fragmentation and results in another traffic jam of dust in this inner region. 

To compute the ionization balance for grains of various size distribution, 
we use the semi-analytical ionization model of \citet{2009ApJ...698.1122O}.
Our ionizing sources include the short-lived radionuclide $^{26}{\rm Al}$ \citep{2009ApJ...690...69U} and stellar X-rays \citep{1999ApJ...518..848I,2009ApJ...701..737B} with an X-ray luminosity of $10^{31}~\rm erg~s^{-1}$.
The radial profile of the ionization rate per hydrogen molecule at the midplane, $\zeta$, is plotted in Figure~\ref{fig:dust}(c).
At $R \ga 15~\rm au$, X-rays are able to penetrate down to the midplane and produce a high value of $\zeta$. 
All positive ions are represented by ${\rm HCO}^+$, with its recombination rate coefficient given by $2.4\times 10^{-7}(T/300~\rm K)^{-0.69}~\rm cm^{3}~s^{-1}$ \citep{1988PhRvA..37.2543G}.
Only for dust evolution and ionization calculations, we assume the gas disk model $\Sigma_g = M_{\rm disk}\exp(-R/R_c)/(2\pi R_c R)$ and $T = 310(R/1~\rm au)^{-0.57}$ with $M_{\rm disk} = 0.2M_\sun$ and $R_c = 150~\rm au$ \citep{2016ApJ...821...82O}, 
where $R$ is the radial distance. The electron and ion abundances obtained from our ionization calculation are shown in  Figure~\ref{fig:dust}(d). 
Reduction of the electron and ion abundances by dust grains is visible in the regions where small dust particles pile up (for comparison, $n_e/n$ ($=n_i/n$) in the dust-free case is also shown in Figure~\ref{fig:dust}(d)).
At $R \ga 15~\rm au$, the effect of dust grains is relatively minor because of the elevated ionization rate.\footnote{It can be shown analytically that $n_e, n_i \propto \zeta$ in the limit where grain charging dominates over gas-phase recombination, 
and $n_e, n_i \propto \zeta^{1/2}$ in the opposite, dust-free limit \citep[e.g.,][]{2009ApJ...698.1122O}.
These indicate that the relative importance of grain charging decreases as $\zeta$ increases. }

\begin{figure}
\centering
\includegraphics[width=0.48\textwidth]{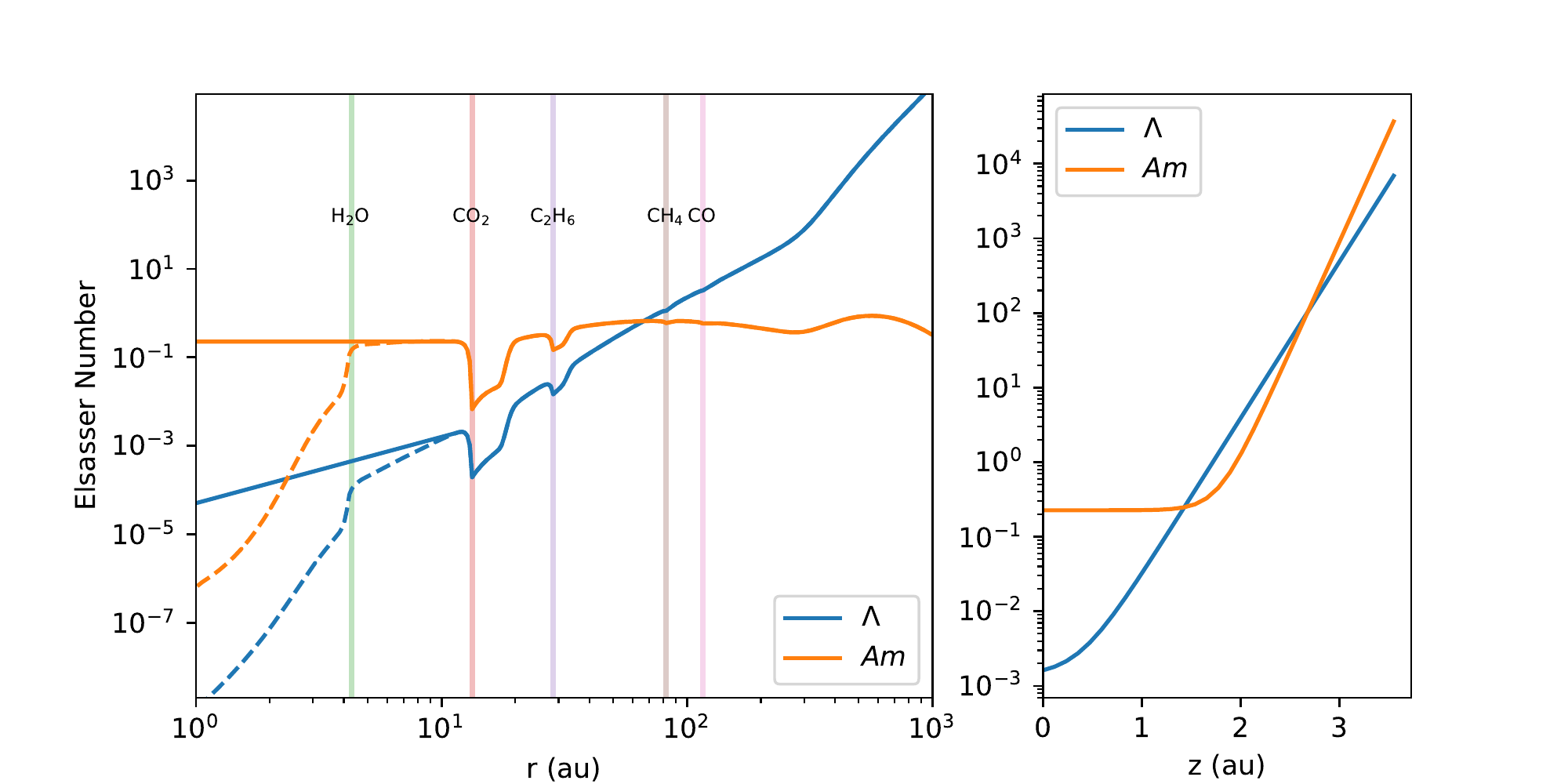}
\caption{
Radial and vertical profile of Elsasser numbers of Ohmic resistivity ($\Lambda=v_a^2/\left(\eta_\mathrm{O}\Omega_K\right)$) and ambipolar diffusion ($Am=v_a^2/\left(\eta_\mathrm{A}\Omega_K\right)$) at disk midplane ( assuming constant plasma $\beta=2\times10^4$ at midplane). In the radial direction, from inside out are $\mathrm{H_2O}$, $\mathrm{CO_2}$, $\mathrm{C_2H_6}$,$\mathrm{CH_4}$ and $\mathrm{CO}$ snow lines. The diffusion profile produced by chemical calculation are plotted in dashed lines, and the solid lines are implemented in our MHD simulation. Note we removed any feature inside $\mathrm{CO_2}$ snow line. The analytical functions of vertical diffusion profiles are described in Equation \ref{eq:diff_z}.
}
\label{fig:eta}
\centering
\end{figure}

The second step is to convert the midplane electron and ion abundances into the midplane Ohmic and ambipolar diffusivities, $\eta_O(R,0)$ and $\eta_A(R,0)$. 
 We use the exact expressions of the diffusivities derived from the conductivity tensor \citep[e.g., Equations~(13) and (15) of][]{2008MNRAS.388.1223S} to compute $\eta_O(R,0)$ and $\eta_A(R,0)$ for $B = 1~\rm mG$. While  $\eta_O$ is independent of $B$, $\eta_A$ depends on $B$. However, as long as $n_i = n_e$, the ambipolar diffusivity is simply proportional to $B^2$ \citep{2001ApJ...552..235B}. Because the condition $n_i \approx n_e$ holds in most part of the disk at $R \ga 4~\rm au$ (see Figure~\ref{fig:dust}(d)), we assume $\eta_A \propto B^2$ when we compute $\eta_A(R,0)$ for $B \not= 1~\rm mG$.   
Because of the $B^2$ dependence, 

the dimensionless ambipolar Elsasser number, $Am=v_a^2/\left(\eta_\mathrm{A}\Omega_K\right)$
is better suited to describe the strength of ambipolar diffusion. Here $v_a=B/\sqrt{4\pi\rho}$ is the Alfven velocity, $\Omega$ is orbital angular velocity. 

The obtained diffusivities are shown in the left panel of Figure~\ref{fig:eta} by the dashed lines. The diffusion gets exceptionally strong inside water snow line. To ensure a reasonable timestep, we focus on the $\mathrm{CO_2}$ and $\mathrm{C_2H_6}$ snow lines, and removes any feature within water snow line. The solid lines are actual profiles we used in our MHD simulations.

The final step is to extend the midplane diffusivities to the diffusivities at arbitrary $z$.
The vertical variation of gas ionization state is more complicated when considering both thermal dynamics and chemical evolution \citep{2017ApJ...847...11W}. For simplicity we adopt an analytical profile similar as \citet{2017ApJ...836...46B}. The strength of both diffusion effects drops quickly when getting away from disk portion, the difference is $Am$ remains constant within $2h$, and $\eta_O$'s decrease starts at midplane.
\begin{eqnarray}
\eta_O\left(R,z\right)&=&\eta_O(R,0)(0.59(1-\tanh(\frac{|z|/10h-0.06}{0.07}))\\ 
Am(R,z)&=&\frac{Am\left(R,0\right)}{0.5-0.5\tanh(20(|z|/10h-0.3))}
\label{eq:diff_z}
\end{eqnarray}
Here $h$ is disk scale height at midplane. These profiles ensure strongest diffusion in the disk midplane and drops gradually over the atmosphere. The AD Elsasser number $Am$ is almost constant till $\approx 2h$ above disk midplane, making AD is more dominant over Ohmic resistivity at disk surface, where density is lower.
\subsection{disk setup}
To minimize grid noise of the radially flowing disk wind, we perform our simulation in spherical polar coordinates $(r,\theta,\phi)$. In this paper, we transform some quantities to cylindrical coordinates $(R,\phi,z)$ for analysis. Note that $R=1$ equals 10 $au$ in physical unit.
For an accretion disk with the vertical hydrostatic equilibrium, the vertical density profile can be determined from temperature. We divided disk vertically into two parts: the high density region within $\approx 2h$ above midplane (disk portion, defined later), and the hotter low density atmosphere. In between we use a sinusoidal function to smooth out the temperature transition. The overall profile is similar to \citet{2017ApJ...836...46B}, but we vary temperature on $z/h$ instead of $\theta$, and the actual value at disk portion and atmosphere is close to \citet{2010MNRAS.401.1415O}, with central star mass extended from 0.8 $M_\odot$ to 1 $M_\odot$:

\begin{equation}
T(R,z)=
\begin{cases}
T(R,0) & \text{if }  |z| < 2h \\
T(R,0)+49.5T(R,0)\times\\
 \ \ (1+\sin(\frac{|z|/h-2}{10}\pi-\frac{\pi}{2})) & \text{if } 2h \leq |z| \leq 12h  \\
100T(R,0); & \text{if } |z| > 12h\\
\end{cases}\label{eq:T}
\end{equation}
The radial grid in our simulation is set from r = 0.1 to 100 in code units with logarithmic grid spacing. The $\theta$ grid extends from 0.05 to $\pi-0.05$.The azimuthal $\phi$ grids spans from $0$ to $\pi/2$. The initial density and temperature profile at midplane both follow power law function with index marked as $p$ and $q$, respectively. Midplane density at $R=1$ is set to be unity in code unit.
The initial disk profile for density and temperature at midplane is:
\begin{eqnarray}
\rho(R,z=0)&=&\rho(R_0,z=0)\left(\frac{R}{R_0}\right)^p\\
T(R,z=0)&=&T(R_0,z=0)\left(\frac{R}{R_0}\right)^q
\end{eqnarray}
The initial vertical density profile is calculated by assuming hydrostatic equilibrium at $R-z$ plane, i.e,$v_R=v_z=0$.
Following \citet{2013MNRAS.435.2610N}, we use the force balance equations at vertical and radial direction to determine $v_\phi$ on $R-z$ plane.
The gas density in the atmosphere can be orders of magnitude lower than the midplane. In order to maintain a reasonable timestep size, a density floor that varies with location is applied (similar to \citet{2018ApJ...857...34Z}, with minor modification close to the pole region):
\begin{equation}
\rho_{fl}(R,z)=
\begin{cases}
\rho_{0,fl} (\frac{r_{min}}{R_0})^p(\frac{r_{min}}{z})^2\frac{r_{min}}{\sqrt{R^2+z^2}} & \text{if }  R \leq r_{min} \\
\rho_{0,fl} (\frac{R}{R_0})^p(\frac{r_{min}}{z})^2\frac{r_{min}}{\sqrt{R^2+z^2}} & \text{if }  R > r_{min} \\
\end{cases}\label{eq:drag}
\end{equation}
where $r_{min}$ is the radius of spherical inner boundary. This density floor follows same power law slope ($p$) as the initial density profile at midplane, and gets smaller at disk atmosphere. We choose $\rho_{0,fl}=3\times 10^{-7}$ in code unit. 

The initial magnetic field is set to be pure vertical. The initialization of magnetic field is described by vector potential $\bm{A}$ so that $\div B=0$, same as \citet{2018ApJ...857...34Z}. The actual field strength is:
\begin{equation}
B(R,z)=Bz(R)=
\begin{cases}
B_0(\frac{r_{min}}{R_0})^{\frac{p+q}{2}} & \text{if }  R \leq r_{min} \\
B_0(\frac{R}{R_0})^{\frac{p+q}{2}} & \text{if }  R > r_{min} \\
\end{cases}
\end{equation}
so we have $B^2\propto R^{p+q}$, the same as midplane gas pressure, to maintain a constant plasma $\beta$ at midplane.

The actual disk parameters are set to match the disk model used in magnetic diffusivity calculation, as p=-2.2218, q=-0.57 \citep[see also,][]{2016ApJ...821...82O}. Unlike that model, we do not apply exponential taper for disk density at the outer region to avoid very small density at the outer boundary. We use the same magnetic field strength as \citet{2017ApJ...845...31H}, i.e, a constant plasma $\beta=2\times10^{4}$ at disk midplane at initialization. Following \citet{2016ApJ...821...82O}, the disk's aspect ratio $H/R$ is 0.058 at $R=R_0$. The simulation domain has $128\times 64\times 16$ grid cells in $r,\theta,\phi$ directions at the root level. The grid size in $\phi$ direction is twice the value of the other two directions, as the initial setup is axisymmetric and we mainly focus on radial and vertical structure evolution. We use three levels of static mesh refinement towards the midplane, so the disk scale height at r=1 is resolved by about 9 grids at the finest level. We use open boundary condition in $r$ direction while restricting mass flux from outside into the domain. Reflective and periodical boundaries are used for $\theta$ and $\phi$ direction, respectively.

\section{Results}\label{sec:results}
\begin{figure}
\centering
\includegraphics[width=0.5\textwidth]
{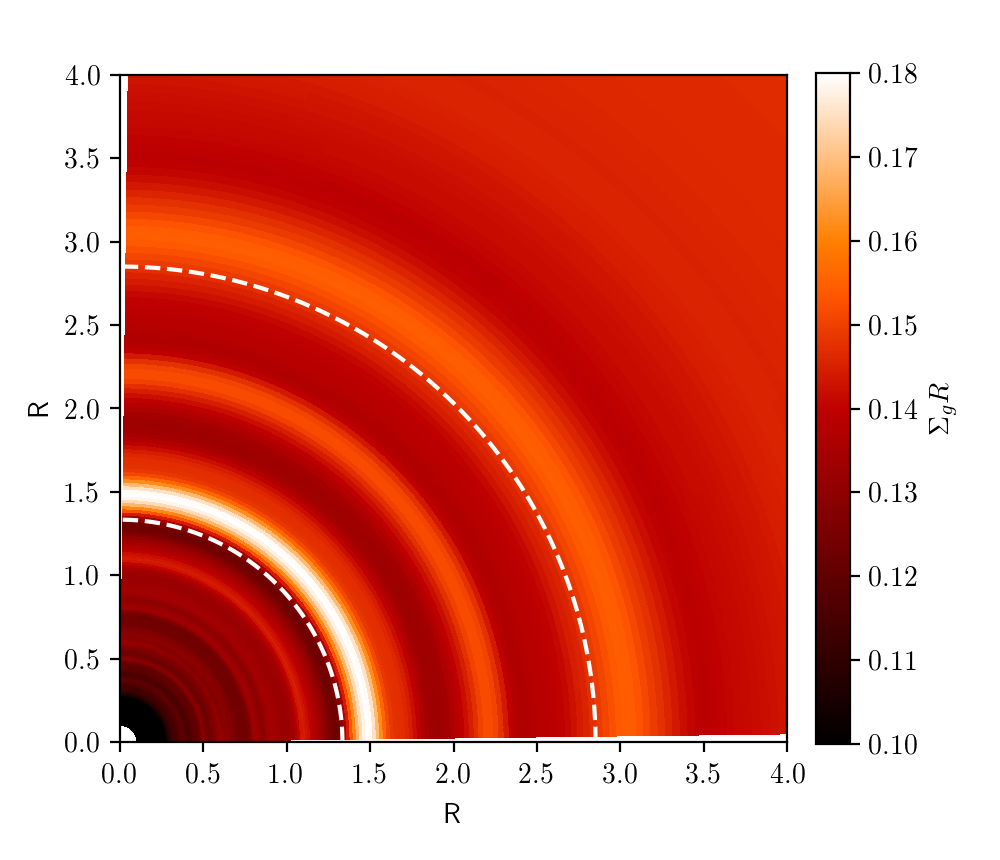}
\caption{2D surface density structure in code units, plotted as $\Sigma_g R$, at $t=50T_0$. The local minima of Elsasser numbers are marked as white dashed lines, correspond to the $\mathrm{CO_2}$ and $\mathrm{C_2H_6}$ snow lines, respectively. Note the "extra" ring between those two snow lines. 
An animation is available. The video begins at $t=0.0$ and ends at $t=50.0$. The real time duration of the vide is 25 seconds.
}
\centering
\label{fig:dens2d}
\end{figure}
We run our simulation for t=50$T_0$, with $T_0$ as one orbital period at r=1 ( 10 au in real world). This is equivalent of 1581 orbits at inner boundary. At this stage, our disk reaches quasi-steady state within R$\sim$ 3, so we can evaluate disk structure within the  $\mathrm{CO_2}$ and $\mathrm{C_2H_6}$ snow lines ($<$ 30 au). The outer disk cannot reach steady state as we do not run sufficient long time and no material is fed at the outer boundary. 
The first thing we notice from our simulations is rings and gaps in gas surface density. In Figure \ref{fig:dens2d} 
we see axisymmetric gaps overlay with the jumps of non-ideal MHD diffusion coefficients that are caused by $\mathrm{CO_2}$ and $\mathrm{C_2H_6}$ snow lines. 
We define the "disk portion" as the region below the wind launching location (where $v_\theta$ rises significantly from zero), which is roughly 0.13 radians above the midplane at r=1.0, slightly above two scale heights. The value gets higher to larger radii as the disk is flared, but for our analysis we'll keep this value constant as our mostly interested region is between r=1 and r=3. 

The most apparent structure is the gap/ring at $r\sim1.3$ ($\mathrm{CO_2}$ snow line). Comparing to a smooth disk profile, the density contrast $\approx 17\%$. This contrast is lower than $37\% \sim 78\%$ contrast we saw in surface brightness \citep{2015ApJ...806L...7Z}. Considering dust is usually more concentrated at pressure bump, the actual dust density profile could be closer to observations.

\subsection{Radial accretion variation}
\begin{figure}
\centering
\includegraphics[width=0.5\textwidth]{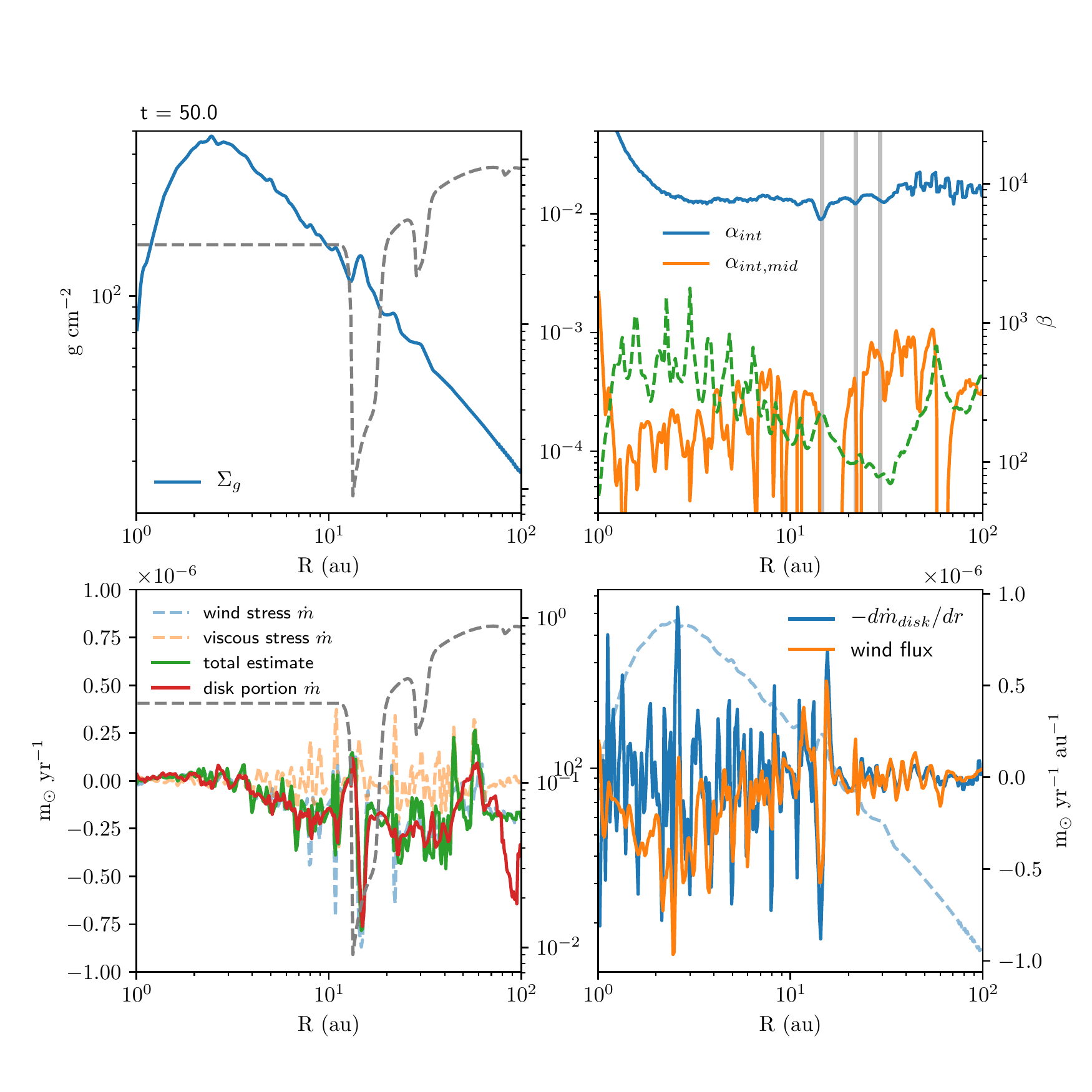}
\caption{
Radial profiles ( from 1 to 100 au) of azimuthally and time average quantities (from t=49 to t=50 $T_0$ with $\Delta t = 0.1$) over radius, with $Am$ overplotted as grey dashed line on right y-axis on left two panels. Top left: gas surface density $\Sigma_g$, in $\mathrm{g\ cm^{-2}}$. 
Top right: vertically integrated $\alpha$, with blue line as total value, and orange line only contains the disk portion. The green dashed line is vertically averaged $\beta$ within the disk portion, shown on right y-axis. The three grey vertical lines marks location of location $\Sigma_g$ maxima ("rings") measured in the top left panel. Bottom left: measured $\dot{m}$ versus calculation from stress, with red solid line as measured midplane $\dot{m}$. The blue dashed line is $\dot{m}$ calculated from wind stress, orange dashed line is $\dot{m}$ calculated from viscous stress, and the green solid line is the sum of these two. Bottom right panel shows quantities measured in spherical polar coordinates. The blue dashed line is $\Sigma_g$ but integrated in $\theta$ direction within the disk portion (in $\mathrm{g\ cm^{-2}}$, y-axis on the left). The blue solid line the radial gradient of mass flux in $r$ direction, within the disk portion, and the orange solid line is local wind flux perpendicular to the disk surface ( wind base). These two quantities are marked by right y-axis, in $\mathrm{\dot{m}\ yr^{-1}\ au^{-1}}$. Note that in this plot, positive $-d\dot{m}_{disk}/dr$ means disk gaines mass at this radii, while positive wind flux means disk loses mass.}
\label{fig:stressmdot}
\centering
\end{figure}

For a viscous accretion disk, the gas radial velocity is $v_{r,g} = -3 \nu  /(2r\sqrt{1-r_*/r})$ \citep{2002apa..book.....F}, i.e, gas drifts radially faster with a higher viscosity. If we calculate the Shakura-Sunyaev
$\alpha$, defined as the $R-\phi$ stress to pressure ratio:
\begin{equation}
\alpha_{int}=\left.\int_{z-}^{z+}{\bm T}_{R\phi}dz  \middle/\int_{z-}^{z+}P_{\rm gas}dz\right.
\end{equation}
where ${\bm T}=\rho v_R \left(v_\phi-\bar{v_\phi}\right)-B_RB_\phi$, $z-$ and $z+$ are upper and lower boundary of the simulation domain. From top right panel of Figure \ref{fig:stressmdot} we show $\alpha_{int}$, the $\alpha$ drops are consistent with $Am$ dips. As gas flows inward, the $v_{r,g}$ drops quickly so gas piles up, creating axisymmetric rings; then the viscosity jumps up very sharply within the $\alpha$ dip, just like the $Am$ profile, so $v_{r,g}$ increases and drains the material behind the ring, then a gap is formed.

\begin{figure*}
\centering
\includegraphics[width=1.0\textwidth]{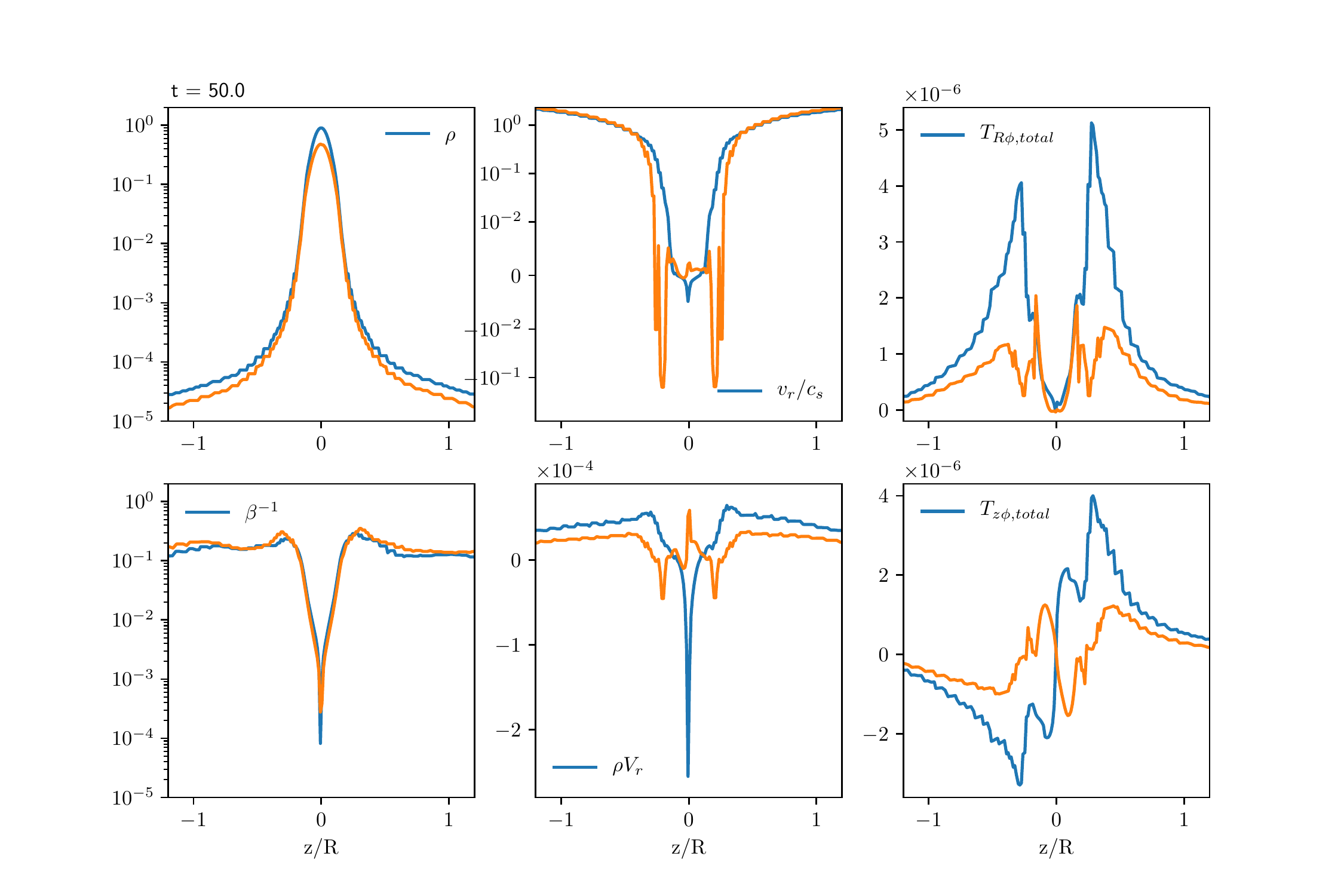}
\caption{
Various quantities along the vertical direction at R=1 (blue) and R=1.33 ( orange, local diffusion maximum at $\mathrm{CO_2}$ snow line). The quantities are averaged over both the azimuthal direction and time(from t=49 to t=50 $T_0$ with $\Delta t = 0.1T_0$). Top row from left to right: gas density, gas radial velocity to local sound speed ratio,$R-\phi$ stress; bottom row: inverse of plasma $\beta$, radial mass flux and $z-\phi$ stress. }
\label{fig:zcutmulti}
\centering
\end{figure*}
In a magnetically coupled disk-wind system, more field diffusivity makes the disk more laminar, and wind can play a key role in driving accretion. Angular momentum transport through wind generated from disk surface can be characterized by the  $z-\phi$ stress:

\begin{equation}
{\bm T}_{z\phi}=\left(\rho v_z \left(v_\phi-\bar{v_\phi}\right)-B_zB_\phi\right)
\end{equation}
The total accretion rate is driven by radial gradient of $R-\phi$ stress and the difference of $z-\phi$ stress between
upper and lower disk surface:
\begin{equation}
\frac{\dot{m}_{acc}v_K}{4\pi}=\frac{d}{dr}\left(R^2\int_{\rm bot}^{\rm up}{\bm T}_{R\phi}dz\right)+R^2\Big|_{\rm bot}^{\rm up}{\bm T}_{z\phi}
\label{eq:stress}
\end{equation}
Here $up$ and $bot$ are upper and bottom wind base. The measured accretion rate is plotted in solid red line in the bottom left panel of Figure \ref{fig:stressmdot}. $\dot{m}$ that is calculated from stress based on Equation \ref{eq:stress} is shown in solid green line, which matches reasonably well with direct measurement. The drop and rise shape of $\dot{m}$ matches what is given by the $\alpha$ profile.

Note the accretion driven by wind stress is much more dominant over the viscous stress in the midplane. Viscous stress $\dot{m}$ is swinging around zero across the disk. We can also justify this by plotting $\alpha$ that only considered the disk portion (Figure \ref{fig:stressmdot} upper right, orange line). The picture that non-ideal MHD diffusion suppresses MRI at midplane and accretion is driven by wind is consistent with previous study \citep{2013ApJ...769...76B}.

A more direct point of view on the evolution of $\Sigma_g$ is measuring the change of the mass accretion rate ($-d\dot{m}/dr$) at each radius. This analysis is done in spherical polar coordinates so we can get the wind flux ($\rho v_\theta$) perpendicular to defined disk surface. The accretion rate is calculated by integrating $\rho v_r$ instead of $\rho v_R$. From the bottom right panel of Figure \ref{fig:stressmdot} we find the local wind loss and the radial change of $\dot{m_{disk}}$ show same scale of contribution to $\Sigma_g$. Since in our calculation, the mass of disk mesh at $r$ grows with positive $-d\dot{m}_{disk}/dr$ while declines with positive wind flux, the overlapping lines in this panel means the disk $\Sigma_g$ is at quasi-steady state, as mass transfer between accretion and wind almost cancels out each other at $t=50T_0$.

Figure \ref{fig:stressmdot} also raises the question of total mass budget. Modeling of previous millimeter data
suggest an HL Tau disk mass of $M_d \approx 0.1 M_\odot$ \citep{2015ApJ...808L...3A,2015ApJ...808..102K}, with age about 1 to 2 Myr. In our simulation, the mass accretion rate is in a reasonable range, $\approx 2\times 10^{-7}m_\odot \ yr^{-1}$. The caveat is wind loss rate. By integrating "wind flux" in bottom right panel of Figure\ref{fig:stressmdot}, we find the total wind loss rate can be up to $\approx 5.5\times10^{-6} m_\odot/yr$, this is more than an order of magnitude over accretion rate. This particularly high wind flux could be due to the overly dense disk atmosphere. 

\subsection{Inside midplane: magnetic reconnection}
To understand the accretion beneath the wind launching point, we plot the vertical cut of azimuthally averaged quantities at R=1 (blue lines) and R=1.33 (orange lines, location of $\mathrm{CO_2}$ snow line) in Figure \ref{fig:zcutmulti}. In the top left panel we see gas density peaks at midplane and drops sharply within the disk portion. Beyond the disk the decrease is slower as temperature gets higher. The density profile at R=1 and at $\mathrm{CO_2}$ snow line do not show significant difference. Same is for the magnetic pressure to thermal pressure ratio, plasma $\beta^{-1}$, at lower left panel. The upside down spike of $\beta^{-1}$ at midplane indicates strong field stretch. The top middle panel shows gas radial ( in spherical $r$ ) velocity over local sound speed ( $v_r/c_s$ ). At R=1.0, the accretion is concentrated at midplane. Wind starts at $\theta-\pi/2=0.13$ above the midplane, and became supersonic at $\theta-\pi/2=0.57$. With similar $v_r$ structure in the wind region, $v_r$ in disk portion is quite different at R=1.33. At midplane we see decretion, and fast ( up to 0.1 local sound speed) accretion happening at disk surface, which is similar to flow pattern in global ideal MHD simulations \citet{2018ApJ...857...34Z}. The similar trend can be seen in radial mass flux ($\rho v_r$) plot at the the lower middle panel. The $R\phi$ and $z\phi$ components of total stress is plotted in rightmost two panels. The $R\phi$ stress at both radius are close to zero at midplane, as already shown by very low midplane $\alpha$ in Figure \ref{fig:stressmdot}. Almost zero $R\phi$ stress at different radius also suggesting that midplane accretion or decretion is not driven by $R\phi$ component. As calculated in Section \ref{sec:results}, vertical ($z$) gradient of ${\bm T}_{z\phi}$ also contributes to $\dot{m}$. In the $z\phi$ stress panel, $d{\bm T}_{z\phi}/dz$ maximizes at midplane, which is in agreement with the $\rho v_r$ spike in the lower middle panel. At R=1.33 (13.3 au), the midplane decretion is driven by $z\phi$ stress's negative $z$ gradient. The surface accretion is related to the two positive $z$ gradient parts around disk surface. The differences of ${\bm T}_{z\phi}$ at these two radius also suggest very different magnetic field structure.

\begin{figure*}
\centering
\includegraphics[width=1.0\textwidth]{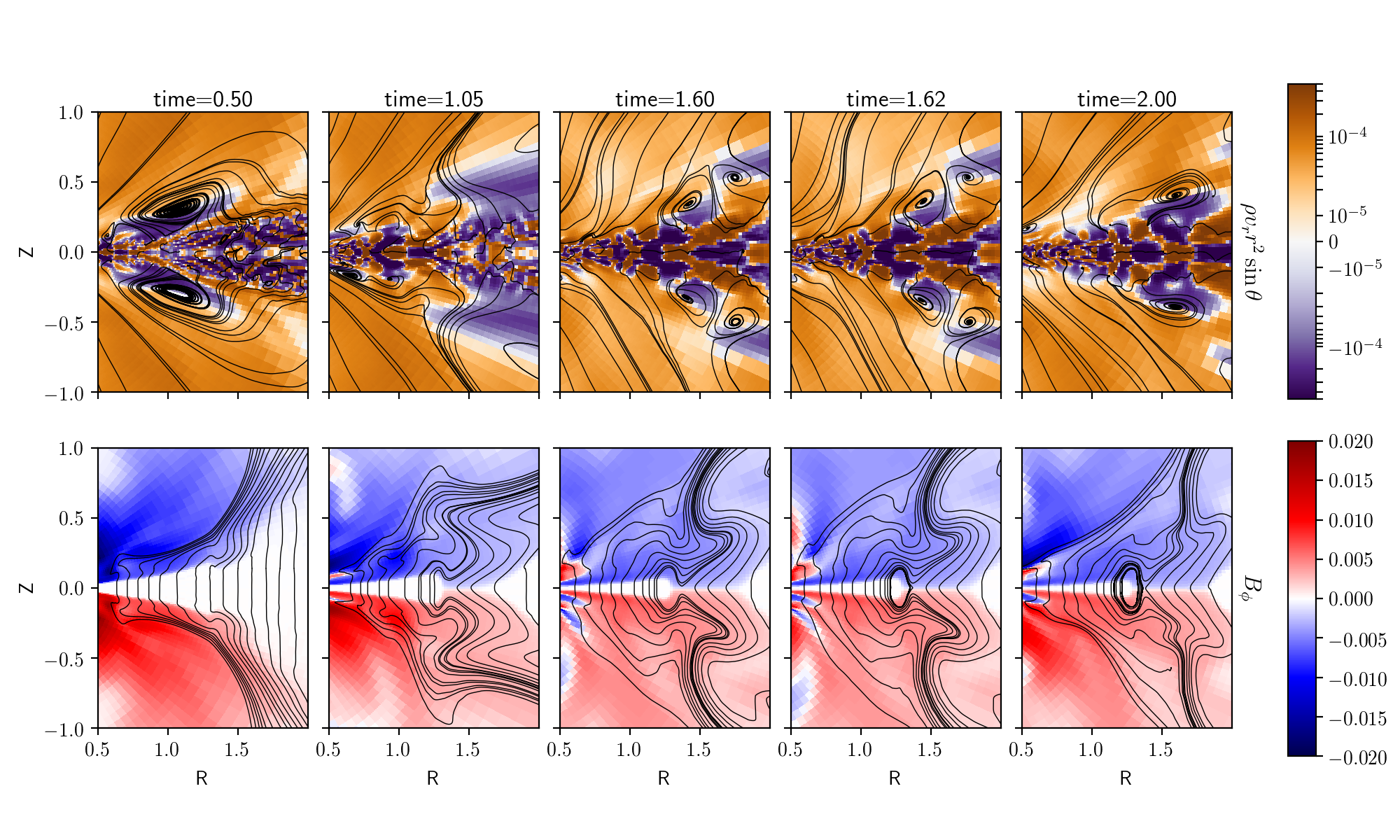}
\caption{
Time sequence of disk evolution at early phase. The upper panels are effective flux ($\rho v_r r^2 \sin \theta$),  and bottom panels are $B_\phi$. Velocity lines and field lines are over-plotted on each panel. The time interval was chosen unevenly to better illustrate the process of magnetic field reconnection at midplane. Note how the form of field lines being varied just above disk surface with transient fast accretion.}
\label{fig:reconnect}
\centering
\end{figure*}

The origin of the differences in stress structure can be understood from the 2D plot on $z-R$ plane, showing how magnetic field is evolved in Figure \ref{fig:reconnect}. At the beginning of simulation, the magnetic field only has vertical ($z$) component, so $B_\phi, B_R=0$ everywhere. Apart from the dominant Keplerian shear ($dv_\phi/dR$) at midplane, the disk is also differentially rotating vertically ($dv_\phi/dz$), as midplane being the fastest, and the atmosphere portion is mostly pressure gradient supported, rotates much slower. After simulation starts, the field lines closer to midplane are dragged towards the rotation direction, making the $B_\phi$ positive below midplane and negative above midplane. The motion along the $r$ direction is even more complicated. From t=0.5, at $\mathrm{CO_2}$ snow line, the velocity shear between wind and surface accretion forms a vortex like flow pattern, that drives gas infall to the midplane. This gas motion drags field lines towards midplane, making a kinked form of magnetic field. Accretion at midplane also drags field lines inward, amplifying horizontal field component $B_R$, and the polarity of $B_R$ changes sign in the accretion stream. Close to midplane, Keplerian shear stretches horizontal field lines, just like the growth of azimuthal fields in the linear phase of MRI. The process makes the field kink at midplane even sharper. The vertical gradient of $B_\phi$ component tends to get steeper with AD \citep{1994ApJ...427L..91B}.
The strength of AD is proportional to $B^2$, so it minimizes at magnetic null (for $B_\phi$ component, and $B_r$, $B_\theta$ varies slowly in $z$ direction at midplane) where magnetic field changes sign at midplane, stopping AD from smoothing out the sharp field structure.

\begin{figure}
\centering
\includegraphics[width=0.48\textwidth]
{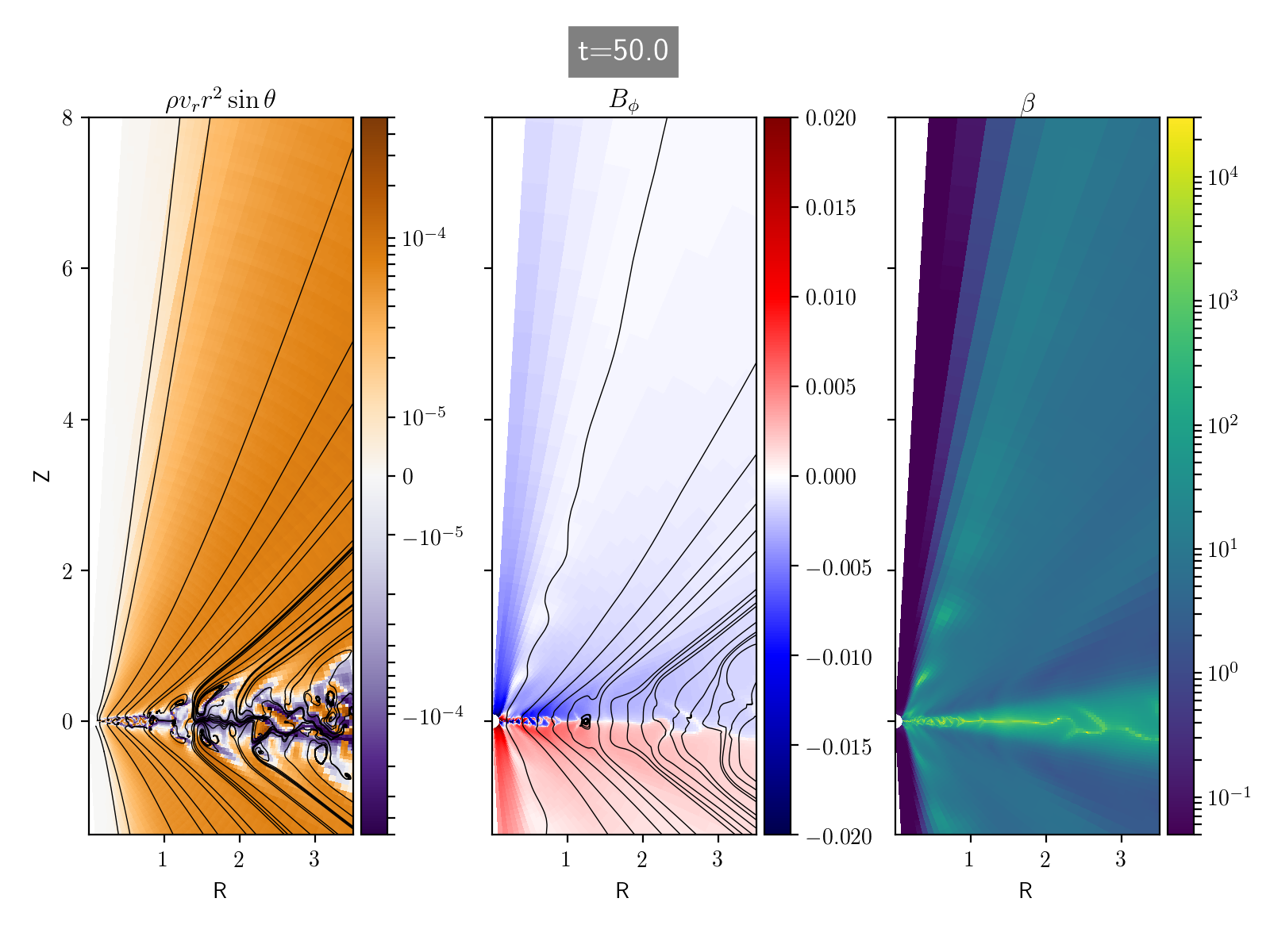}
\caption{The azimuthally  and time averaged (from t=49 to t=50 $T_0$ with $\Delta t = 0.1T_0$) effective mass flux ($\rho v_r r^2 \sin \theta$, left), $B_\phi$ (middle) and plasma $\beta$ (right) at t=50$T_0$. The black curves on the left panel are the velocity field lines calculated with
azimuthally averaged velocities, on the middle panel are field lines calculated with azimuthally averaged magnetic field.
An animation is available. The animation is slightly different from the static figure in that the static figure is an average of 10 outputs while the video shows the evolution sequence from $t=0$ to $t=50.0$. The color map is reversed in the left panel of the animation. The real time duration of the video is 25 seconds.
}
\centering
\label{fig:rhob2d}
\end{figure}

The steep magnetic gradient leads to large $dT_{z\phi}/dz$ at midplane, drives a highly concentrated  accretion sheet. An alternative physical picture is the the removal of angular momentum at midplane by Lorentz force. Since $J_r \sim \partial (B_\phi \sin\theta)/\partial \theta$,  and $B_\phi$ changes sign at midplane, there is a infinite thin current sheet at the disk midplane. The Lorentz force $\propto J_rB_\theta$ then maximizes at midplane and drives accretion at its highest rate. The accretion flow now drags field lines inward, pinching the magnetic field radially. At R=1.33 (13.3 au, $\mathrm{CO_2}$ snow line), the kinked field lines, together with field lines dragged by infalling gas towards midplane, eventually reconnect. The poloidal component $B_\theta$, or vertical component $B_z$ that reconnects now have their direction reversed. This changes sign of  $dT_{z\phi}/dz$ at midplane, driving mass flux outward. This outward moving mass can also be seen in the mass flux panel (left) in Figure \ref{fig:rhob2d}. The outward mass flux concentrates gas just outside the field reconnection area, forming a gaseous ring.

The middle panel of Figure \ref{fig:rhob2d} also shows that the magnetic field does not always changes its polarity exactly at midplane. From R=2.2 to R=2.8, there is an large triangular shaped structure of $B_\phi$ anomaly. $B_\phi$ usually changes from positive to negative when moving from below to above midplane, but at the anomaly, the sequence of field polarity is positive-negative-positive-negative. This changes direction of ${\bm T}_{z\phi}$ thus alters the direction of mass flux. This zone of decretion covers a relatively large area close to midplane, while the outward mass flow in the field loop of magnetic reconnection ($\mathrm{CO_2}$ snow line) is confined in a thin sheet, as plotted in left panel in Figure \ref{fig:rhob2d}. The formation of this magnetic anomaly is shown in Figure \ref{fig:reconnect}. In this area, there's a quick accretion flow above the disk surface. As mentioned before, $B_R$ changes sign in accretion stream , with Keplerian shear, $B_\phi$ also changes sign. This overall $B_\phi$ structure sustains over time, though surface accretion is just a transient behavior in the early phase.

\section{Parameter study}
\label{sec:parameter}
In order to assess how robust the ring formation mechanism in the reference simulation is, especially how the diffusion structure contribute to ring build up, we performed several additional simulations. The first one is a disk with "featureless" diffusion profile, i.e, we use constant $Am$ and $\eta_\mathrm{O}$ at midplane for the whole disk. The exact values are set to match the diffusion strength at R=1.16 in the reference run, with $Am=0.23$ and $\eta_\mathrm{O}=2.2\times10^{15} cm^2s^{-1}$ ($1.56\times10^{-5} $ in code units).

\begin{figure}
\centering
\includegraphics[width=0.48\textwidth]
{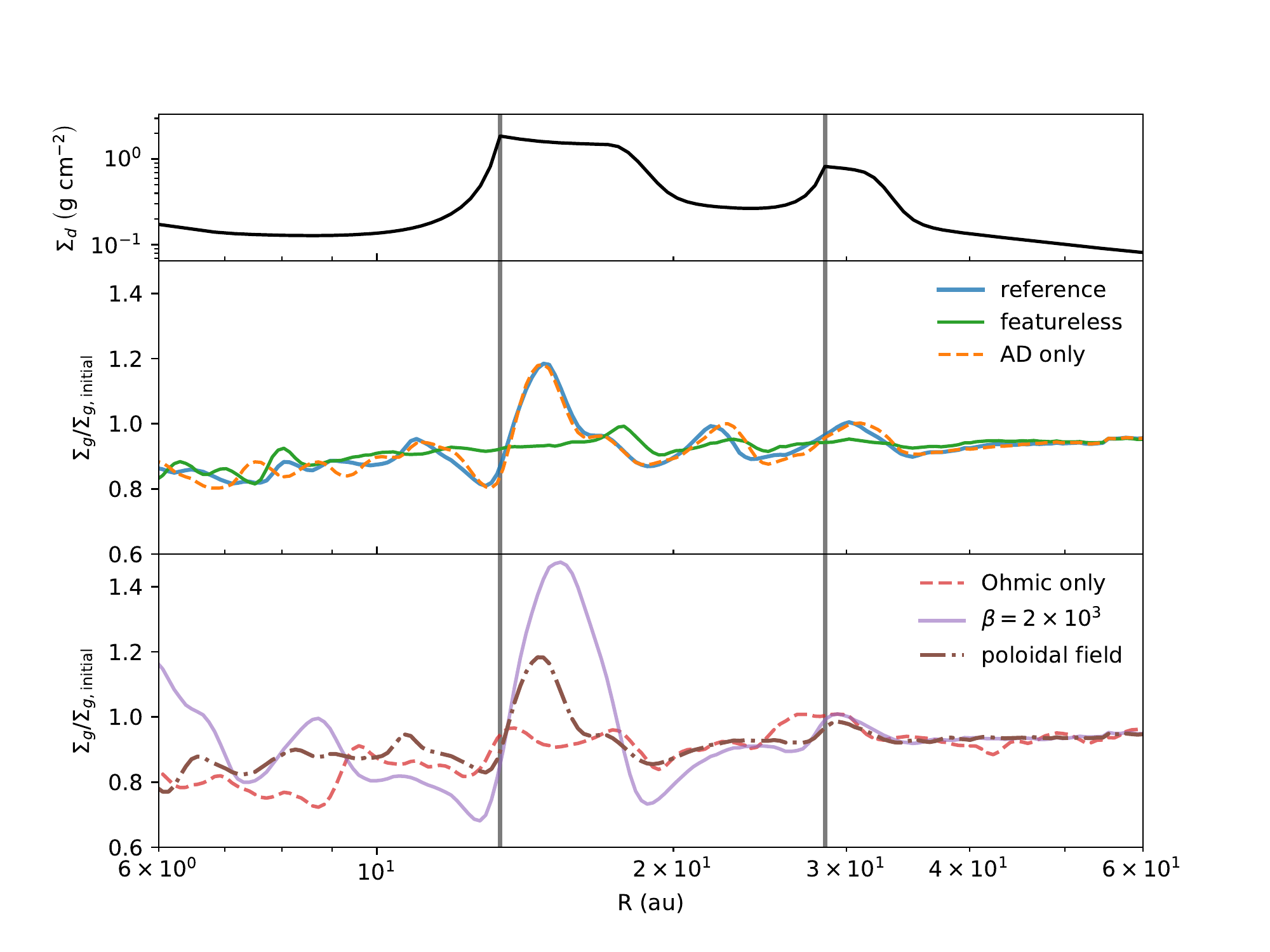}
\caption{Gas surface density variation ($\Sigma_g/\Sigma_{g,\rm initial}$) of six different models, divided in two bottom panels for better illustration. The top panel shows dust surface density adopted for ionization calculation.The vertical lines mark location of $\mathrm{CO_2}$ and $\mathrm{C_2H_6}$ snow line. Note the "extra" density bump is not featured in cases of "poloidal field" and "$\beta=2\times10^3$.
Also note the relative location of gaseous rings and dust rings. The deepest point of gap is located a bit inside the snow lines, as the region of decretion caused by magnetic reconnection extends a bit inward.
}
\centering
\label{fig:sigma}
\end{figure}

\begin{figure*}
\centering
\includegraphics[width=1.0\textwidth]
{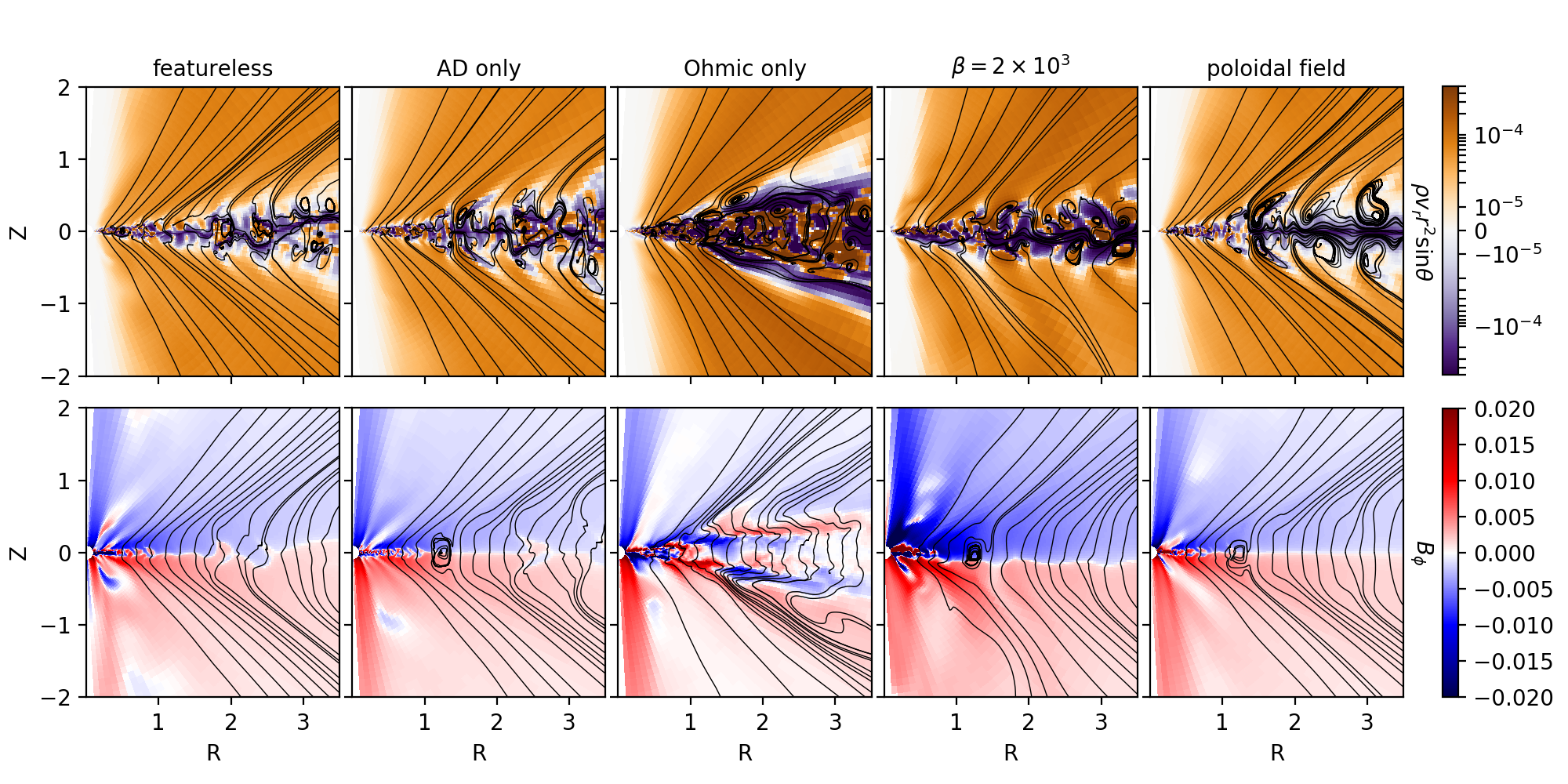}
\caption{The azimuthally averaged effective mass flux ($\rho v_r r^2 \sin \theta$, upper), $B_\phi$ (lower) t=50$T_0$ of five different models. The black curves on the upper panel are the velocity field lines calculated with
azimuthally averaged velocities, on the lower panel are field lines calculated with azimuthally averaged magnetic field. The color scale is the same as in Figure \ref{fig:rhob2d}.}
\centering
\label{fig:rhob2dmulti}
\end{figure*}

The results of "featureless" model are shown as solid green line in Figure \ref{fig:sigma} and leftmost panels in Figure \ref{fig:rhob2dmulti}. The rings and gaps at $\mathrm{CO_2}$ and $\mathrm{C_2H_6}$ snow lines disappeared with smooth diffusion profiles. There is however a density bump at R=1.8. Judging from Figure \ref{fig:rhob2dmulti}, we see very similar $B_\phi$ and mass flow pattern as R=2.2 to R=2.8 in the reference run. This density bump has the same origin as the one at R=2.2 in the reference model. Another significant difference in Figure \ref{fig:rhob2dmulti} is the absence of magnetic reconnection for the featureless run. This is similar to \citet{2018MNRAS.477.1239S,2019MNRAS.484..107S}, as reconnection only happens when AD is strong enough at midplane. The field reconnection in the reference model is very likely to prevent surface accretion from dragging field lines inwards as much as in the featureless run. This moves the $B_\phi$ anomaly outward, thus the "extra" density bump changes from R=1.8 to R=2.2 in the reference model.

The second and third additional simulations are "AD only" and "Ohmic only" runs. We use the same setup as the reference run but only include Ambipolar diffusion or Ohmic resistivity in these simulations. 
The "AD only" model is almost identical to the reference model in both 1D $\Sigma_g$ profile and 2D field and flow pattern. The "Ohmic only" case shows barely any significant feature in disk surface density and a vastly different structure in Figure \ref{fig:rhob2dmulti}. Accretion is dominated by fast flows at disk surface and disk midplane is filled with chaotic patterns of both accretion and decretion. The majority of accretion not being at midplane makes the diffusion features at snow line give little contribution to the whole disk.

In the 4th test we simply make magnetic field $\sqrt{10}$ times stronger than the reference model, setting plasma $\beta=2\times10^3$. Note we keep $Am$ and $\eta_\mathrm{O}$ unchanged so the strength of magnetic diffusion remains the same as reference model. The result is significant thicker rings and deeper gaps at nearly all locations ( Figure \ref{fig:sigma}). This is expected as magnetic stress is 10 times larger and so its radial and vertical gradients. The "extra" density bump at R=2.2 does not show up in this case. The relevant $B_\phi$ anomaly and flow pattern are also missing in $\beta=2\times10^3$ panels in Figure \ref{fig:rhob2dmulti}. With plasma $\beta$ 10 times smaller, the magnetic field is so strong that the initial surface accretion is not able to drag and bend it to make the $B_\phi$ anomaly.

In the last test, "poloidal field", we do not start with a pure vertical field. Poloidal magnetic fields are initialized with vector potential generalized from \citet{2007A&A...469..811Z}:
\begin{equation}
A_\phi(r, \theta) = \frac{2B_{z0}R_0}{3-\alpha}\left(\frac{R}{R_0}\right)^{-\frac{\alpha-1}{2}}\
[1+(m\tan\theta)^{-2}]^{-\frac{5}{8}}
\end{equation}
where $m$ is a parameter that specifies the degree that poloidal fields bend, with $m$ goes infinity giving a pure vertical field.  We chose $m=0.5$ the same as \citet{2017ApJ...836...46B}. The initially bent field lines follows the gas flow better than vertical field, and is not easily dragged by fast surface inflow when the simulation starts. This setup gets rid of the $B_\phi$ anomaly. In Figure \ref{fig:sigma}, the surface density profile of "poloidal field" almost overlaps with our reference run, with the absence of the "extra" density bump at $R=2.2$. It advises that MHD simulation with moderate AD and magnetic field may better start with a poloidal field instead of pure vertical field. 

\subsection{Observation signature}
Observational tests for the derived disk structure in our simulations include: 1) studying dust and gas surface density  across the gaps (Figure 8), 2) probing the azimuthally averaged gas velocity structure (Figure 7) using the method in \cite{2018ApJ...860L..12T}, 3) measuring the disk magnetic field orientation and strength directly. 

\begin{figure}
\centering
\includegraphics[width=0.48\textwidth]
{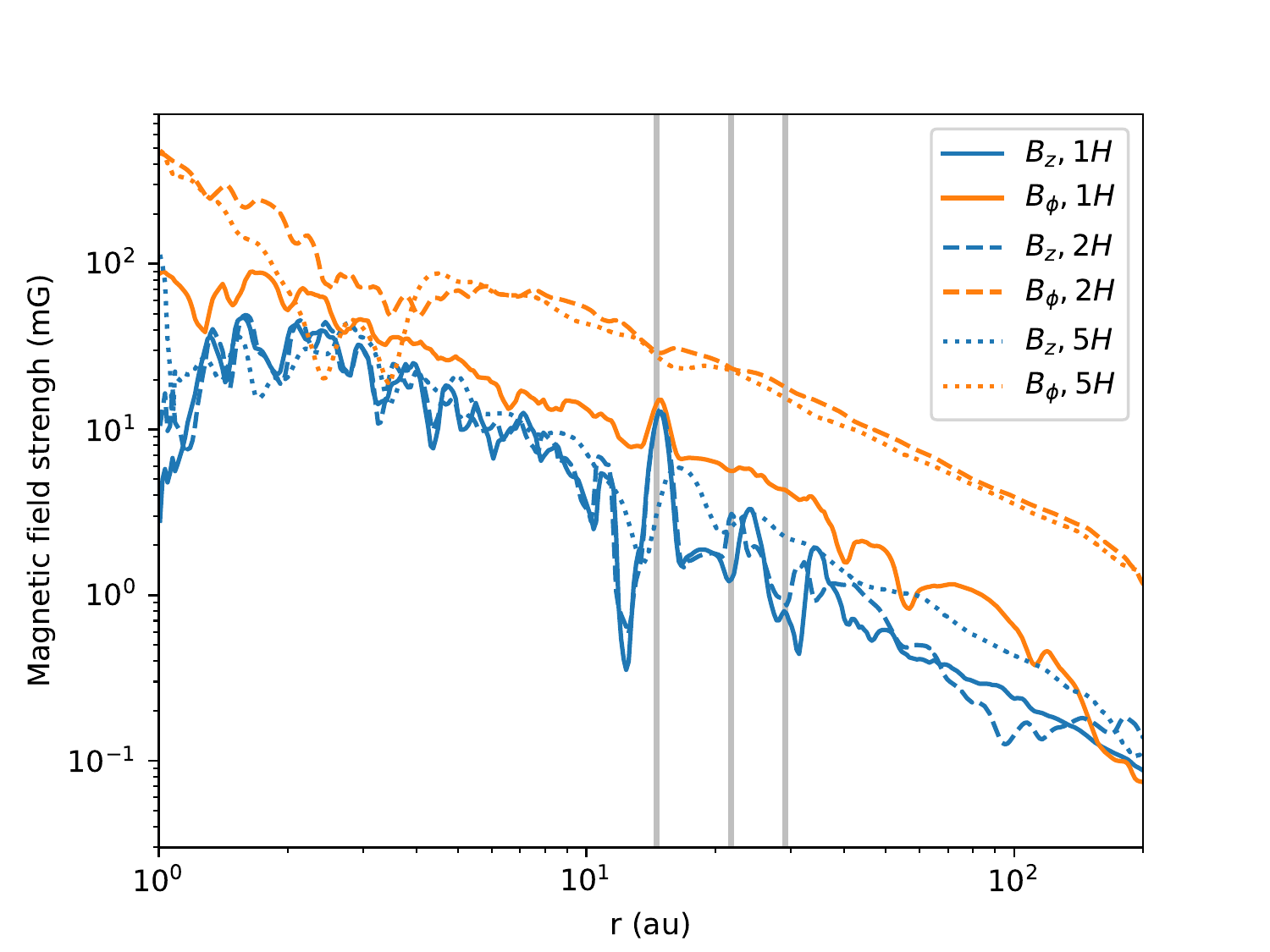}
\caption{Measured magnetic field (poloidal component $B_z$ and toroidal component $B_\phi$) strength of the reference model, at z=h, 2h and 5h above midplane.}
\centering
\label{fig:BzBp}
\end{figure}

One exciting recent observational constraint is from using the molecular lines' Zeeman effect to study disk magnetic fields directly.
Recent ALMA circular polarization measurements for CN lines in TW Hya suggest that, at 42 au, the upper limit of $B_z$ and $B_\phi$ are $0.8\ \mathrm{mG}$ and $30\ \mathrm{mG}$, respectively \citep{2019A&A...624L...7V}. Chemistry modelling shows CN are most abundant at z/R between 0.3 and 0.4 beyond 10 au \citep{2018A&A...609A..93C}. In Figure \ref{fig:BzBp} we measured poloidal (vertical) and toroidal magnetic field at z=h, 2h and 5h above midplane. At r=50 au, z=5h gives $B_z=1\ \mathrm{mG}$ and $B_\phi=8\ \mathrm{mG}$, which are roughly consistent with the derived upper limit. This also suggests that ALMA Zeeman measurements start to constrain MHD disk models. The future circular polarization measurements for molecular lines will provide
stringent tests to disk accretion models.

We have shown the field strength in our reference run. To scale the field strength to systems with other accretion rate, we measured $\dot{m}$ of $\beta=2\times10^3$ case. It is $\approx 1.5\times 10^{-6}m_\odot \ yr^{-1}$, 7 times higher than the reference run. This gives $\dot{M}\propto|B|^{1.7}$, and it is between the wind driven accretion measurement from \citet{2013ApJ...772...96B} (almost proportioanl to $|B_z|$), and the MRI driven accretion in \citet{2014ApJ...785..127O} (proportional to $B_z^2$).

\section{Summary}
\label{sec:sum}

In this study, we have carried out a set of 3D non-ideal MHD simulations of magnetically coupled disk-wind systems with different profiles of ambipolar diffusion and ohmic resistivity. Our main conclusions are as follows.

(i) snow line in HL Tau make dust aggregates pile up in the sintering zones due to the
combined effect of radial drift and sintering-induced fragmentation. The concentration of smaller dust leads to stronger ambipolar diffusion and Ohmic resistivity in these areas. By varying magnetic diffusion coefficients we can generate gaseous rings and gaps in a weakly ionized gas disk with moderate magnetic field. 
Generally speaking, varying diffusion strength changes accretion structure, making material pile up radially. This is a new dust feedback mechanism. Comparing to aerodynamic drag, it requires much less dust to take effect.

(ii) The correlation between snowline locations and rings/gaps can be complex. The dust model we adopted is based on smooth gas disk, while the rings produced in our simulations can trap dust and change the overall profile of $\Sigma_d$. In Figure \ref{fig:sigma} the snowlines are located a bit outside the deepest point in the gap. The width of gaseous rings in the reference run is overall narrower than rings in the dust model. As dust are being trapped at pressure maxima, the radius with strongest diffusion (also $\Sigma_d$ maxima) can move outward, towards the gaseous ring. If the zone of magnetic reconnection relocates at the pressure maxima, it reduces both dust and gas surface density, slows down ring formation. The exact strength of this gas to dust back reaction still needs further investigation. One factor that complicates this scenario is the ones are most concentrated at pressure bumps due to radial drift are larger dust particles, while in ionization calculation it is the smallest grains that dominates the total electron absorbing cross section. 

(iii)With very strong ambipolar diffusion ($Am\sim0.01$), the gaseous rings and gaps can be generated from magnetic reconnection at midplane. High AD strength confines accretion at midplane and the vertical differentiation of radial velocity drags field lines so close that they reconnect. The reconnected field loop alters the direction of midplane mass flow. This drains mass inside the reconnection zone and carves gaps in the disk. The mass accumulated outside the field loop form dense rings.

(iv) Compared to ambipolar diffusion, Ohmic resistivity is almost irrelevant to ring formation at outer part of disk. In the "Ohmic only" case, the disk is better magnetically coupled and behaves more like ideal MHD. The accretion is dominated by mass flow at disk surface instead of midplane in the AD simulation. One contribution factor could be that Ohmic resistivity is only strong at high density region. It declines rapidly from midplane to disk surface. A probably more important distinction between Ohmic resistivity and ambipolar diffusion is the former does not rely on the mutual angle between current and magnetic field, so the diffusion from Ohmic is isotropic. Ohmic resistivity is less effective in modifying magnetic field and accretion structure.

(v) With moderate magnetic field strength ($\beta=2\times10^4$), AD dominated non-ideal MHD simulation can be sensitive to initial profile of magnetic field. Fast surface flow drags field lines radially, and then the radial pinch is twisted by vertically differentiated rotation. Eventually this process leaves a $B_\phi$ anomaly at midplane, and a circular like flow pattern is sustained. This flow pattern allows the formation of an "extra" ring between the $\mathrm{CO_2}$ and $\mathrm{C_2H_6}$ snow line. Stronger magnetic field will not be affected by initial surface velocity shear. A poloidal field that is bent away from rotation axis follows the fast surface flow better than pure vertical field, and is also barely affected.  We'd like to point out the fact that the "extra" ring is related to specific initial setup does not rule out this ring formation mechanism in real astrophysical process. Formation of certain rings may not be related to specific diffusion structure. 

(vi) The wind loss rate in our simulations is exceptionally high. The vertical structure of the disk, including both temperature and non-ideal diffusion, are simple prescribed profiles.  \citet{2019ApJ...874...90W} presented first MHD disk simulation with consistent thermochemistry. Temperature at disk atmosphere and magnetic diffusivities can be obtained directly in real time. This would be a nice direction of improvement for furture study.

Overall, our work suggests that changes in the disk ionization structure can robustly lead to the change of the disk accretion structure, which leads to gaseous gaps and rings. In our particular example, the change of the dust size distribution across
the snowline affects the non-ideal MHD effects and eventually leads to gaseous gaps and rings.
Our work also suggests that, besides these changes of the disk ionization structure, non-ideal MHD effects in smooth disks can also lead to structure formation but these structures sensitively depends on the numerical setup and different non-ideal MHD effects involved. 

\section*{Acknowledgements}

We thank Mordecai Mac-Low, Phil Armitage, Jeremy Goodman and Zhenyu Wang for helpful discussions, and Jiming Shi for help on the code. X. H. and Z. Z. acknowledge support from the National Aeronautics and Space Administration through the Astrophysics Theory Program with Grant No. NNX17AK40G and Sloan Research Fellowship. Simulations are carried out with the support from the Texas Advanced Computing Center (TACC) at The University of Texas at Austin through XSEDE grant TG-AST130002. S.O. is supported by JSPS KAKENHI Grant Numbers JP16K17661, JP18H05438, and JP19K03926. X.-N.B acknowledges support from the Youth Thousand Talent Program of China.K.T. is supported by JSPS grants JP16H05998, JP16K13786 and JP17KK0091. K.T. acknowledges support by the Ministry of Education, Culture, Sports,
Science and Technology (MEXT) of Japan as Exploratory Challenge on Post-K 
computer (Elucidation of the Birth of Exoplanets [Second Earth] and the 
Environmental Variations of Planets in the Solar System).

\end{document}